\documentclass[prl,twocolumn,superscriptaddress,amsmath,amssymb]{revtex4-1}
\usepackage{graphicx}

\begin{document}

    \title{Linear-in-frequency optical conductivity in GdPtBi due to transitions near \\ the triple points}
    \author{F. H\"{u}tt}
    \affiliation{1.~Physikalisches Institut, Universit\"at Stuttgart, 70569 Stuttgart, Germany}
    \author{A. Yaresko}
    \affiliation{Max-Planck-Institut f\"ur Festk\"orperforschung, 70569 Stuttgart, Germany}
    \author{M. B. Schilling}
    \affiliation{1.~Physikalisches Institut, Universit\"at Stuttgart, 70569 Stuttgart, Germany}
    \author{C. Shekhar}
    \author{C. Felser}
    \affiliation{Max-Planck-Institut f\"{u}r Chemische Physik fester Stoffe, 01187 Dresden, Germany}
    \author{M. Dressel}
    \author{A. V. Pronin}
    \affiliation{1.~Physikalisches Institut, Universit\"at Stuttgart, 70569 Stuttgart, Germany}
    \date{October 23, 2018}

\begin{abstract}

The complex optical conductivity of the half-Heusler compound GdPtBi
is measured in a frequency range from 20 to 22 000~cm$^{-1}$ (2.5
meV -- 2.73 eV) at temperatures down to 10 K in zero magnetic field.
We find the real part of the conductivity, $\sigma_{1}(\omega)$, to
be almost perfectly linear in frequency over a broad range from 50
to 800 cm$^{-1}$ ($\sim$~6 -- 100 meV) for $T \leq 50$ K. This
linearity strongly suggests the presence of three-dimensional linear
electronic bands with band crossings (nodes) near the chemical
potential. Band-structure calculations show the presence of triple
points, where one doubly degenerate and one nondegenerate band cross
each other in close vicinity of the chemical potential. From a
comparison of our data with the optical conductivity computed from
the band structure, we conclude that the observed nearly linear
$\sigma_{1}(\omega)$ originates as a cumulative effect from all the
transitions near the triple points.

\end{abstract}

\maketitle

Heusler materials are currently well recognized for their wide range
of spectacular electronic and magnetic
properties~\cite{Wollmann2017}. The high tunability of these
compounds allows designing materials with properties on demand for
future functioning applications~\cite{Manna2018, Casper2012}.
Recently, band inversion and topologically nontrivial electronic
states have been intensively studied in (half-)Heusler compounds
with strong spin-orbit coupling (SOC) \cite{Chadov2010, Lin2010,
Xiao2010, Al-Sawai2010, Liu2011, Liu2016, Logan2016, Shekhar2016}.

Among other half-Heusler compounds with strong SOC, GdPtBi occupies
a special place: the hallmarks of a Weyl-semimetal (WSM) state, such
as negative magnetoresistance and the planar Hall effect, are
vividly developed in this material and are assigned to
manifestations of the chiral anomaly~\cite{Hirschberger2016,
Liang2018, Kumar2017}. It has been proposed that the band structure
of GdPtBi in zero magnetic field can be sketched as two degenerate
parabolic bands touching each other at the $\Gamma$ point of the
Brillouin zone (BZ) \cite{Hirschberger2016}. A moderate external
magnetic field splits the bands. This leads to linear-band crossings
and a WSM state, which enables negative longitudinal
magnetoresistance~\cite{Hirschberger2016, Liang2018} and the planar
Hall effect~\cite{Kumar2017}. Most recent density-functional-theory
calculations, though, forecast linear-band crossings even in zero
magnetic field~\cite{Yang2017}. These nodes are, however, different
from the Dirac and Weyl points: in the GdPtBi case, one doubly and
one nondegenerate band cross each other, forming the so-called
triple points~\cite{Zhu2017}. Angular-resolved photoemission reveals
linear electronic bands in GdPtBi, but these bands are mostly
assigned to the surface states~\cite{Liu2011}.

Because of large penetration depths, optical methods are more
sensitive to bulk properties~\cite{Dressel2002}. It is also known
that optical transitions between bands with linear dispersion
relations manifest themselves as characteristic features in the
optical response~\cite{Ando2002, Hosur2012, Bacsi2013, Carbotte2017,
Mukherjee2017, Ahn2017}. For example, crossing three-dimensional
(3D) linear bands are supposed to show up as linear-in-frequency
conductivity, $\sigma_1(\omega) \propto \omega$. Such linearity of
$\sigma_1(\omega)$ --~unusual for conventional metals and
semiconductors~-- is currently widely considered as a hallmark for
solid-state 3D Dirac physics and has indeed been observed in a
number of nodal semimetals~\cite{Chen2015, Xu2016, Neubauer2016,
Kimura2017, Shao2018}. Therefore, it is tempting to probe the
optical response of GdPtBi and to compare it with theory
predictions.

In this paper, we report on measurements of the optical conductivity
in GdPtBi. We find $\sigma_1(\omega)$ to be linear in a broad
frequency range: at $T \leq 50$~K, the linearity spans from $\sim
100$ meV down to a few meV. We calculate $\sigma_1(\omega)$ from the
GdPtBi band structure and, by comparing the experimental and the
computed conductivity, demonstrate that the linear-in-frequency
$\sigma_1(\omega)$ is due to electronic transitions between the
bands in the vicinity of the triple points.

GdPtBi single crystals were grown by the solution method from a Bi
flux. Freshly polished pieces of Gd, Pt, and Bi, each of purity
larger than 99.99 $\%$, in the ratio Gd:Pt:Bi =1:1:9 were placed in
a tantalum crucible and sealed in a dry quartz ampoule under 3 mbar
partial pressure of argon. The filled ampoule was heated at a rate
of 100 K/hr up to 1200$^{\circ}$C, followed by 12 hours of soaking
at this temperature. For crystal growth, the temperature was slowly
reduced by 2 K/hr to 600$^{\circ}$C. Extra Bi flux was removed by
decanting it from the ampoule at 600$^{\circ}$C. Overall, the
crystal-growth procedure followed closely the ones described in
Refs.~\cite{Shekhar2016, Canfield1991}. The crystals' composition
and structure (noncentrosymmetric $F\overline{4}3m$ space group)
were checked by energy dispersive x-ray analysis and Laue
diffraction, respectively.

\begin{figure}[b]
\centering
\includegraphics[width=0.9\columnwidth]{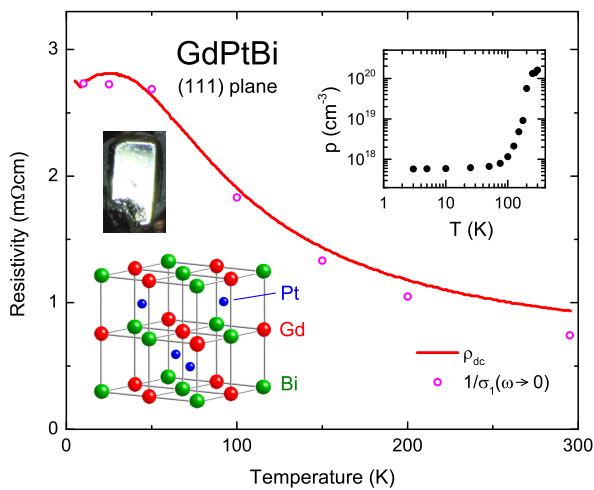}
\caption{Temperature-dependent dc resistivity (solid line) of GdPtBi
and the inverse values of its optical conductivity at $\omega
\rightarrow 0$ (open dots). Inset: carrier concentration $p$
obtained from Hall measurements. The sample used in this work is
also shown alongside the GdPtBi structure.} \label{res}
\end{figure}

Our optical reflectivity measurements were conducted on a single
crystal of lateral dimensions of $\sim 2 \times 1.1$ mm$^{2}$ with a
shiny (111) surface (Fig.~\ref{res}); the sample thickness was
around 0.8 mm. Standard four-point dc-resistivity and Hall
measurements, performed on a smaller piece (a Hall bar) cut from the
specimen used for optics, indicated a semiconducting behavior with a
well-known antiferromagnetic transition at 9 K \cite{Canfield1991}.
Hall measurements show $p$-type conduction and a very low carrier
density of $6 \times 10^{-17}$ cm$^{-3}$ at $T \rightarrow 0$ (cf.
different samples from Ref.~\cite{Hirschberger2016}). All optical
experiments reported here are made in the paramagnetic state ($T
\geq 10$ K), where the Dirac physics of GdPtBi is primarily
discussed~\cite{Hirschberger2016, Liang2018, Kumar2017, Yang2017}.

Optical reflectivity $R(\nu)$ was measured at 10 to 300 K over a
frequency range from $\nu = \omega/(2\pi c) = 20$ to $22 000$
cm$^{-1}$ (2.5 meV -- 2.73 eV) using two Fourier-transform infrared
spectrometers. The spectra in the far-infrared ($20 - 700$
cm$^{-1}$) were recorded by a Bruker IFS 113v spectrometer. Here, an
\textit{in-situ} gold evaporation technique was utilized for
reference measurements~\cite{Homes1993}. At frequencies above 700
cm$^{-1}$, a Bruker Hyperion microscope attached to a Bruker Vertex
80v spectrometer was used. Freshly evaporated gold mirrors served as
references in this setup. Complex optical conductivity was obtained
from $R(\nu)$ using Kramers-Kronig
transformations~\cite{Dressel2002}. High-frequency extrapolations
were made utilizing the x-ray atomic scattering
functions~\cite{Tanner2015}. At low frequencies, we used the same
procedure as recently applied for materials with highly mobile
carriers~\cite{Schilling2017Yb, Schilling2017Zr}: we fitted the
$R(\nu)$ spectra with a set of Lorentzians~\cite{fit} and then used
the results of these fits between $\nu = 0$ and 20 cm$^{-1}$ as
zero-frequency extrapolations for subsequent Kramers-Kronig
transformations.

\begin{figure}[t]
\centering
\includegraphics[width=\columnwidth]{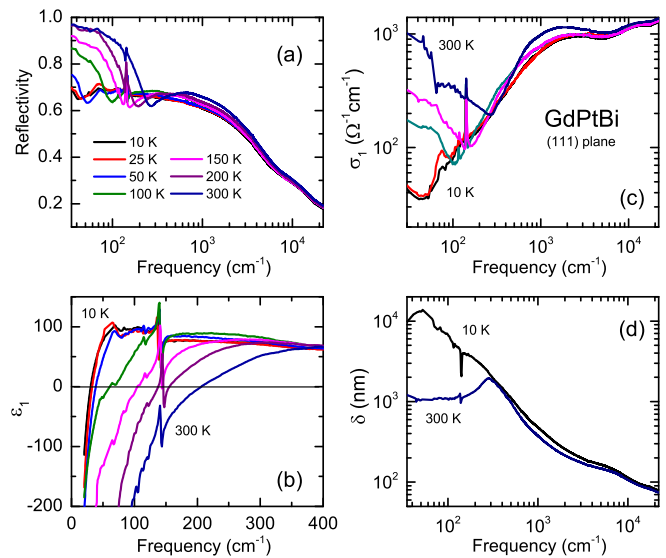}
\caption{Reflectivity (a), dielectric constant $\varepsilon_1$ (b),
optical conductivity $\sigma_1$ (c), and skin depth $\delta$ (d) of
GdPtBi at different temperatures as indicated. The low-frequency
portion of $\sigma_1(\nu)$ is magnified in Fig.~\ref{lin}}
\label{R_delta}
\end{figure}

Figure~\ref{R_delta} displays an overview of the results obtained in
our optical investigations. Panel (a) shows the reflectivity for all
measurement temperatures. Panels (b) and (c) represent (the real
parts of) the dielectric constant $\varepsilon_1(\nu)$ and the
optical conductivity $\sigma_1(\nu)$, respectively. Finally, panel
(d) demonstrates the skin depth $\delta(\nu)$ of our sample at 10
and 300 K (curves for intermediate temperatures lie in between these
ones). Important is that the skin depth exceeds 50 nm for all
measured temperatures and frequencies. In the most interesting
low-energy region, it is in the micrometer range. Hence, our optical
measurements reflect bulk properties.

A sharp phonon peak is seen in all data sets at $\sim 140$
cm$^{-1}$. Another phonon at $\sim 115$ cm$^{-1}$ is weak, but
resolvable, especially in $\varepsilon_1(\nu)$ [panel (b)]. The
frequency positions of both phonon modes have only marginal
temperature dependence. No other phonons are detected, in agreement
with group analysis, which predicts two infrared-active optical
modes for the half-Heusler structure~\cite{phonons}. All other
features of the optical response are due to intra- or interband
electronic transitions, as discussed below.

A temperature-dependent plasma edge dominates the low-energy part
($\nu < 300$ cm$^{-1}$) of the reflectivity spectra [panel (a)]. The
edge corresponds to the screened plasma frequency of free carriers,
$\nu_{\rm{pl}}^{\rm{scr}}$ \cite{Dressel2002}, and is also seen in
panel (b) as zero crossings of the $\varepsilon_1(\nu)$ curves. From
the same panel, it can also be noted that the background dielectric
constant $\varepsilon_{\infty}$ is rather high, around 70 -- 100.
This leads to a low unscreened plasma frequency $\nu_{\rm{pl}} =
\nu_{\rm{pl}}^{\rm{scr}} \sqrt\varepsilon_{\infty}$ (for example,
$\nu_{\rm{pl}} \approx 300$ cm$^{-1}$ at 10 K), in agreement with
the low free-carrier concentration found in Hall measurements. The
free-electron contribution to the optical conductivity [panel (c)]
is seen as a Drude-like mode at the lowest frequencies. At lower
temperatures, this mode narrows and loses its spectral weight in
accordance with decreasing $\nu_{\rm{pl}}^{\rm{scr}}$ at $T
\rightarrow 0$. As $T \rightarrow 0$ K, only marginal traces of the
free-carrier (intraband) contribution are seen in the recorded
$\sigma_1(\nu)$ spectra: above $\sim 50$ cm$^{-1}$, $\sigma_1(\nu)$
reflects only the interband optical transitions (and the phonons, as
mentioned above).

\begin{figure}[t]
\centering
\includegraphics[width=\columnwidth]{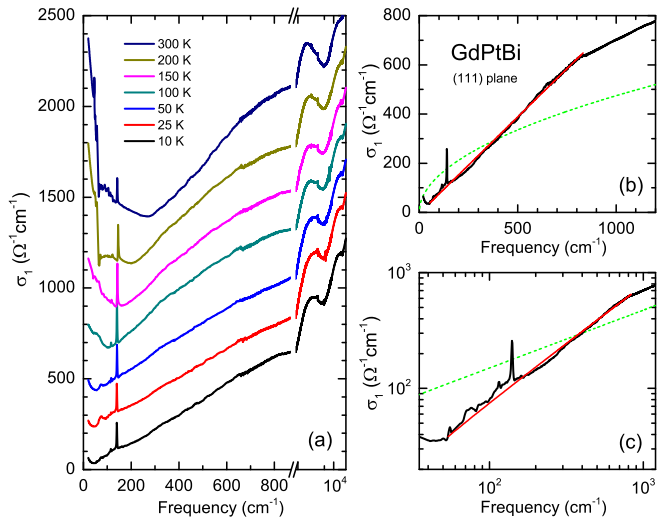}
\caption{Optical conductivity of GdPtBi with an emphasis on low
frequencies (a). Note a frequency-scale change at 900 cm$^{-1}$. The
curves for $T \geq 25$ K are shifted upwards for clarity (by 200
$\Omega^{-1}$cm$^{-1}$ as compered to the previous measurement $T$).
Linear fit (straight red line) of the experimental $\sigma_1(\nu)$
at 10 K for $50 < \nu < 800$ cm$^{-1}$ on linear (b) and log-log (c)
scales. A square-root behavior of $\sigma_1(\nu)$, expected for
parabolic bands, is shown by dashed green lines.} \label{lin}
\end{figure}

A striking feature of the optical conductivity is its almost perfect
linearity in a broad range in the far infrared. This can be seen
best in Fig.~\ref{lin}, where experimental conductivity is shown
alongside linear fits [square-root behavior of conductivity,
expected for parabolic bands, is also shown for comparison]. The
behavior of experimental $\sigma_1(\nu)$ is basically the same for
the three lowest measurement temperatures (10, 25, and 50 K): it
linearly increases with $\nu$ in the spectral range from
approximately 50 up to almost 800 cm$^{-1}$. The observation of this
linearity is an important result of this work. As discussed above,
the linearity of the low-energy $\sigma_1(\nu)$ is a signature of 3D
linear bands~\cite{Hosur2012, Bacsi2013, Chen2015, Xu2016,
Neubauer2016, Kimura2017}. However, other band structures may also
provide similar $\sigma_1(\nu)$. For example, it can be a cumulative
effect of many bands with predominantly, but not exclusively, linear
dispersion relations. Such a situation was recently reported by some
of us for the Weyl semimetal NbP at somewhat higher
frequencies~\cite{Neubauer2018}. Wavy deviations from a perfectly
linear increase of $\sigma_1(\nu)$ [see, Figs.~\ref{lin}(b) and
\ref{lin}(c)] indicate that a similar scenario might be realized in
GdPtBi.

\begin{figure}[]
%\centering
\raggedright
\includegraphics[width=\columnwidth]{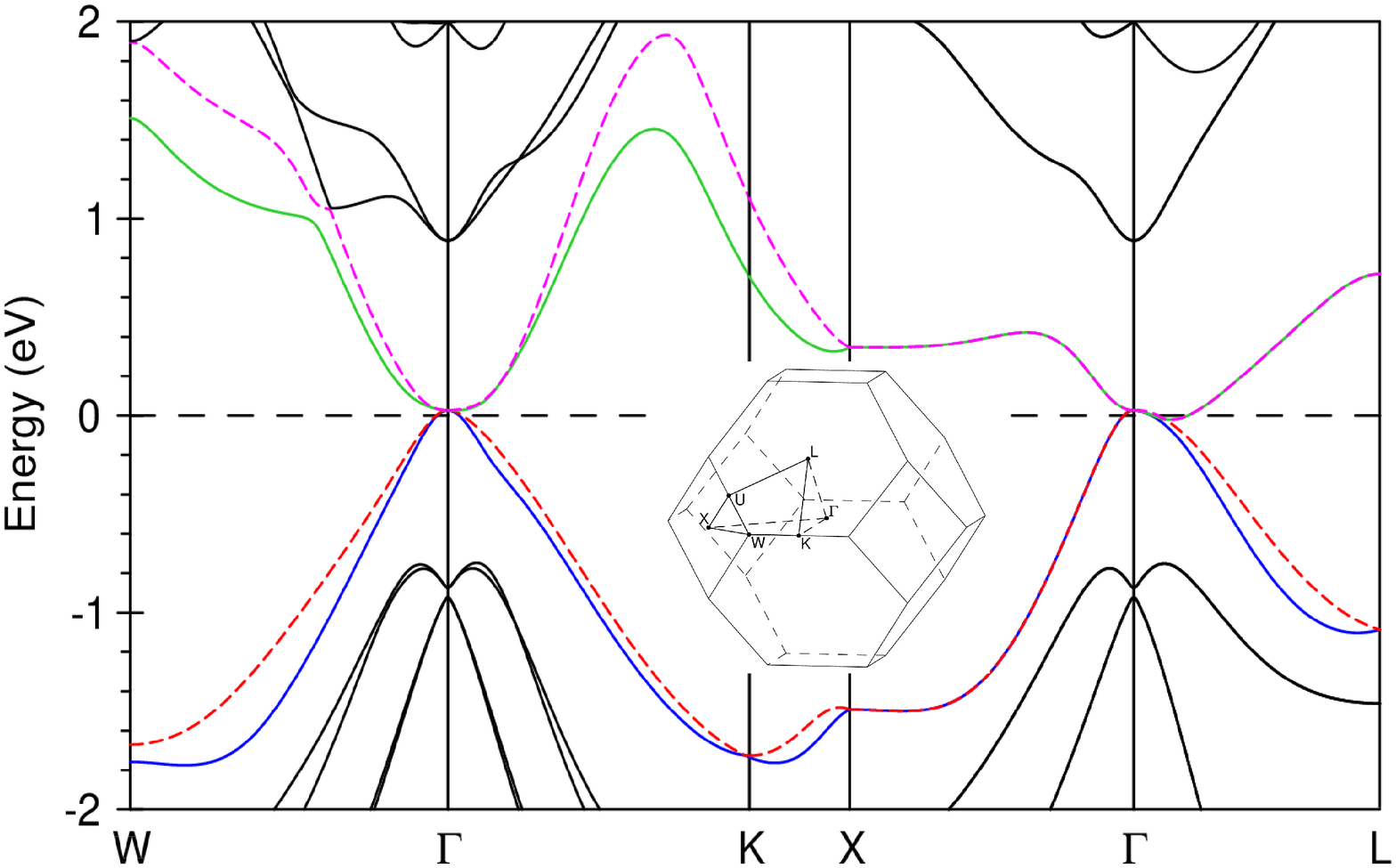}
\caption{Band structure of GdPtBi. Low-energy bands are displayed in
different colors. The BZ is shown as an inset.} \label{bands}
\end{figure}

To get an insight into the origin of linear $\sigma_1(\nu)$, we
performed band-structure calculations and then computed the
interband optical conductivity. In the calculations, we used the
linear muffin-tin orbital method (LMTO) \cite{Andersen75} as
implemented in the relativistic PY LMTO computer
code~\cite{APSY95,AHY04}. The Perdew-Burke-Ernzerhof GGA
\cite{PBE96} was used for the exchange-correlation potential. The
$4f^{7}$ states of gadolinium were treated as semicore states. The
$4f$ spin polarization was not considered in order to model the
paramagnetic state studied in this work ($T \geq 10$ K).
Relativistic effects, including SOC, were accounted for by solving
the four-component Dirac equation inside atomic spheres. BZ
integrations were done using the improved tetrahedron method
\cite{BJA94}.

\begin{figure}[t]
\centering
\includegraphics[width=\columnwidth]{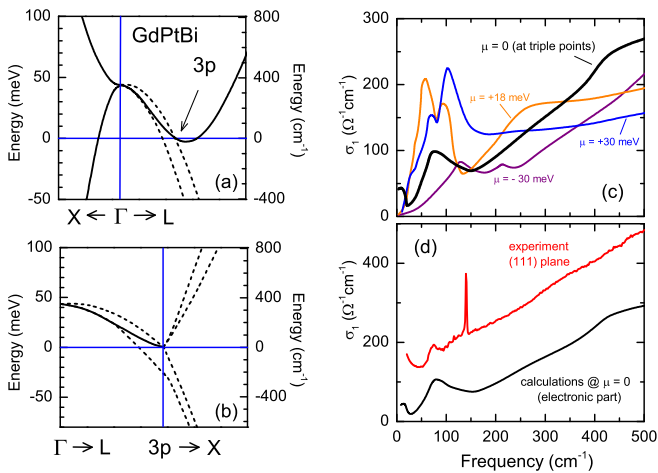}
\caption{Panels (a) and (b): Electronic bands of GdPtBi near the
triple point (marked as 3p). The chemical potential $\mu$ is set to
zero at the triple point. Doubly degenerate bands are shown as solid
lines, while nondegenerate bands as dashed lines. Panel (c):
Calculated interband conductivity of GdPtBi, $\sigma_{1}(\omega)$,
for a few different positions of $\mu$ as indicated. Panel (d):
Comparison of the measured (upper curve) and calculated for $\mu=0$
(bottom curve) optical conductivity of GdPtBi. Phonon modes are not
included in the calculated $\sigma_{1}(\omega)$. The experimental
curve is shifted upwards by 100 $\Omega^{-1}$cm$^{-1}$ for clarity.}
\label{comparison}
\end{figure}

The calculated band structure is shown in Fig.~\ref{bands}. Our
calculations confirm the presence of basically parabolic bands,
touching each other at the $\Gamma$ point. A closer look at the
low-energy band structure [Figs.~\ref{comparison}(a) and
\ref{comparison}(b)] reveals the presence of a triple point (marked
as 3p in the figure) along the $\Gamma - L$ line, in agreement with
previous calculations~\cite{Yang2017}. The bands in the plane normal
to the $\Gamma - L$ direction possess linear dispersion relations,
as shown for the $\rm{3p} - X$ direction in panel (b). There are
eight symmetry-equivalent triple points per BZ. The band structure
of GdPtBi near these points is similar to a Weyl or Dirac semimetal
with tilted cones, as seen best in panel (a).

Our calculations predict the triple points to be situated 18 meV
below the Fermi level. However, the real GdPtBi crystals are known
to often possess unintentional doping, which is impossible to
control at the crystal-growing stage~\cite{Hirschberger2016}. Thus,
the position of the chemical potential can in reality be within a
few tens of meV off the calculated value. It is instructive to note
here that the linear-in-frequency $\sigma_1(\nu)$ is expected for
tilted cones (of any type), if the chemical potential $\mu$ is
situated at the node~\cite{Carbotte2016}. In practice, $2\mu$ ($\mu$
is measured form the node hereafter) should be below the measurement
frequency window. Such a situation can be relevant for our GdPtBi
sample, as we discuss below. This is also in agreement with the very
small free-carrier (Drude) contribution and low Hall carrier
density.

The band structure of GdPtBi in the vicinity of the Fermi level is
obviously more complex than the model band structure used in
Ref.~\cite{Carbotte2016}. Thus, as mentioned above, we compute
$\sigma_1(\nu)$ from the obtained band structure. In these
computations, the dipole matrix elements for interband optical
transitions were calculated on a $96 \times 96 \times 96$ $k$-mesh
using LMTO wave functions~--~it is necessary to use sufficiently
dense meshes in order to resolve transitions at low
energies~\cite{Chaudhuri2017}. The real part of the optical
conductivity was calculated using the Kubo-Greenwood linear-response
expressions \cite{WC74} with the BZ integration performed using the
tetrahedron method.

Before we discuss the results of these calculations, we would like
to note that obtaining a good agreement between experimental and
computed conductivity is known to be challenging~\cite{Yu2010,
Santos2018}. This is particularly the case for (topological)
semimetals, where only a qualitative match can typically be
achieved~\cite{Neubauer2018, Grassano2018, Kimura2017, Frenzel2017,
Chaudhuri2017}. In the relevant for this study low-energy part of
the spectrum (below $\sim 100$ meV), a reasonable agreement is
particularly hard to obtain~\cite{Kimura2017, Chaudhuri2017}.
Nevertheless, for GdPtBi we have reached a fairly good agreement
between our calculations and the experimental spectra at low
energies.

The results are shown in Figs.~\ref{comparison}(c) and
\ref{comparison}(d). Because of the possible carrier doping in
GdPtBi, discussed above, we have some freedom in setting the
position of the chemical potential. We varied $\mu$ within $\pm 30$
meV from the triple point and compared the computed $\sigma_1(\nu)$
spectra to each other and to the experiment. Panel (c) demonstrates
that the best linear $\sigma_1(\nu)$, extrapolating to 0 at $\nu
\rightarrow 0$, is obtained, if the chemical potential is at the
triple point ($\mu = 0$). If we vary $\mu$, the calculated
$\sigma_1(\nu)$ either develops huge peaks at low energies ($\nu <
200$ cm$^{-1}$), or does not extrapolate to 0 as $\nu \rightarrow
0$, or both. Also, the quasilinear part of the conductivity,
calculated for $\mu = 0$, spans over the largest frequency range.
Thus, we choose the $\mu = 0$ curve for further comparison with our
experimental results; see panel (d). (Obviously, very small
deviations from $\mu = 0$ on a meV scale are possible.)

In Fig.~\ref{comparison}(d), a low-temperature (25 K) experimental
curve is shown alongside the calculated $\sigma_1(\nu, \mu = 0)$.
The overall linear increase of the experimental curve is well
reproduced. It is also evident that both the calculations and
experiment provide some deviations from perfect linearity. Most
remarkable is the bump, present in the calculations and experiment,
at around 80 cm$^{-1}$. Such deviations reflect the fact that the
band structure is not ideally linear in all three directions but
more complex. Overall, we can conclude that the observed interband
optical conductivity in GdPtBi originates from the transitions
between all the bands near the triple points. Linear terms dominate
the dispersions of these bands in the close vicinity of the nodes,
leading to the almost, but not perfectly, linear optical
conductivity in GdPtBi at low frequencies.

From our band-structure calculations, we can compute the Fermi
velocities $v_F$ for the crossing bands. Calculations exactly at the
triple point are technically challenging, and, thus, we compute
$v_F$ in a close vicinity of it along the $\Gamma - L$ line -- at
$\pm 0.005 \times 2\pi/a$ from the triple point; here $a$ is the
lattice constant. For the doubly degenerate electronlike band, we
obtain $v_F = 1.1$ and $0.5 \times 10^{5}$ m/s, while for the
nondegenerate holelike band $v_F = 2.4$ and $2.8 \times 10^{5}$ m/s.

In a simple model of electron-hole symmetric crossing linear
electronic bands, the optical conductivity is related to the Fermi
velocity $v_F$ via~\cite{Hosur2012,Bacsi2013}
%\begin{equation}
$\sigma_1(\omega) = \frac{e^2 g N} {24 h} \frac{\omega} {v_F}$,
%\label{interband}
%\end{equation}
where $g$ is the band degeneracy at the crossing point (e.g., a
Dirac node has $g = 4$) and $N$ is the number of nodes per BZ.
Obviously, this simple formula has a very limited applicability.
Nevertheless, if we straightforwardly apply it to our experimental
$\sigma_1(\omega)$ and set $g = 3$ and $N = 8$, we obtain an
averaged Fermi velocity of $\sim 10^{5}$ m/s, which is in good
agreement with the values calculated above.

In summary, we have found the low-frequency optical conductivity of
GdPtBi to be linear in a broad frequency range (50 -- 800 cm$^{-1}$,
$\sim$ 6 -- 100 meV at $T \leq 50$ K). This linearity strongly
suggests the presence of three-dimensional linear electronic bands
with band crossings near the chemical potential. A comparison of our
data with the optical conductivity computed from the band structure
demonstrates that the observed $\sigma_{1}(\omega)$ originates from
the transitions near the triple points. From the optical spectra, we
directly determine the plasma frequency of free carriers in GdPtBi
and estimate an averaged Fermi velocity at the nodes: $v_F \sim
10^5$ m/s. The values of $v_F$, calculated from the band structure
near the triple points along the $\Gamma - L$ line, range from $0.5$
to $2.8 \times 10^5$ m/s depending on the band and the momentum
direction.

We are grateful to Ece Uykur and Dominik G\"{u}nther for valuable
experimental support, to Gabriele Untereiner for technical
assistance, and to Johannes Gooth for many fruitful discussions.
This work was funded by DFG via grant No. DR228/51-1.

\end{document}